\documentclass[fleqn,twoside,11pt]{article}
\usepackage{espcrc2}
\usepackage{latexsym}  % for Symbol fonts
\usepackage{amssymb}
\makeatletter
\def\set@curr@file#1{%
  \begingroup
    \escapechar\m@ne
    \xdef\@curr@file{\expandafter\string\csname #1\endcsname}%
  \endgroup
}
\def\quote@name#1{"\quote@@name#1\@gobble""}
\def\quote@@name#1"{#1\quote@@name}
\def\unquote@name#1{\quote@@name#1\@gobble"}
\makeatother
\usepackage{graphicx}
\usepackage{amsmath}
\usepackage{pstricks}
\usepackage{color}

\begin{document}

% \hfill 2020.03.04
\begin{center}

{\Large\bf Are Charged Leptons in the  \\[0.1in]
Simultaneous Eigenstates  of Mass and Family? }

\vspace{10mm}
{\bf Yoshio Koide}

 {\it ${}^a$ Department of Physics, Osaka University, 
Toyonaka, Osaka 560-0043, Japan} \\
{\it E-mail address: koide@kuno-g.phys.sci.osaka-u.ac.jp}

\end{center}

\begin{quotation}
Conventionally, the observed charged leptons are regarded 
the simultaneous eigenstates of ``mass" and ``family".  
Against this view,  we discuss a possibility that the observed 
charged leptons $e_i=(e, \mu, \tau)$ are not identical with 
the eigenstates of family $e^0_\alpha =(e_1^0, e_2^0, e_3^0)$.  
Here, we define the eigenstates of family, $e^0_\alpha$, 
as the states which interact with family gauge bosons 
in the mass eigenstates of the broken U(3)$_{family}$ gauge
 symmetry.  
Although there is at present not any experimental evidence 
for $e^0_1$-$e^0_2$ mixing, and we have only 
an upper limit for the mixing from the present experimental data. 
We will conclude that the $e$-$\mu$ mixing angle $\theta$ 
must be $\theta \lesssim 10^{-3}$. 
Thus, we can not exclude a possibility $\theta\neq 0$. 
If we want more small upper limit of $\theta$, a rare decay 
search  $\mu \rightarrow e + \gamma$ will be useful.  
\end{quotation}

%%%%%%%%%%%%%%%%%%%%%%%%%%%%%%%%%%%%%%%%%%%%%%%%%%%%
\vspace{5mm}

{\large\bf 1. \ Introduction} \ 

\vspace{2mm}

It is well known that the observed quarks are in the eigenstates
of masses, but they are not in the eigenstates of ``family".
When we denote the eigenstates of mass as $q_i=(u_i, d_i)$ 
($i=1,2,3$), and the eigenstates of family as 
$q^0_\alpha =(u^0_\alpha, d^0_\alpha)$ ($\alpha=1,2,3$),
the relations of both eigenstates are given as follows,
$$
(u_L)_i =(U_u)_i^{\ \alpha} (u_L^0)_\alpha , \ \ \ 
(d_L)_i =(U_d)_i^{\ \alpha} (d_L^0)_\alpha  
\eqno(1.1)
$$
where $(u^0, d^0)_L$ compose SU(2)$_L$ doublets, 
then the Cabibbo-Kobayashi-Maskawa (CKM) 
 mixing matrix \cite{CKM} $U_{CKM}$ is given by
$$
U_{CKM} = (U_u)^\dagger (U_d) .
\eqno(1.2)
$$ 
We know the observed value of the matrix $U_{CKM}$, but, 
at present, we cannot describe the mixing matrices 
$U_u$ and $U_d$ separately, because we do not have 
the definition of the states $(u^0_\alpha, d^0_\alpha)$ 
($\alpha=1,2,3$).
(At present, we know many models where the quark mass matrices 
are given by a non-diagonal form. 
However, the purpose of the present paper is not to discuss 
a mass matrix model which has a special non-diagonal form. )

On the other hand, in the charged lepton sector, 
it is often that  the mass eigenstates 
$(e^-, \mu^-, \tau^-)$ are defined as the eigenstates 
of ``family". 
This definition is simple and clear. 
Against this conventional definition, in this paper, 
we will chose another definition of ``family".
Then, we will be able to see new physics world 
as we discuss in this paper.

\vspace{5mm}

{\large\bf 2. \ Definition of ``family"}

\vspace{2mm}

We introduce family gauge bonsons in a U(3) family symmetry.
We consider a frame where the symmetry is broken, so that
the U(3) family gauge bosons $A_\alpha^{\ \beta}$ acquire   
masses $M_{\alpha\beta}$. 
In such the frame, the eigenstates of family for 
quarks $q^0_\alpha$ are defined as
$$
H_{int} = g_F (\overline{q^0})^\alpha \gamma_\mu (q^0)_\beta 
( A_\alpha^{\ \beta})^\mu.  
\eqno(2.1)
$$
Here, $(q_L)^0_\alpha = ((u_L)^0_\alpha, (d_L)^0_\alpha)$
compose  SU(2)$_L$ doublets. 
Thus, the meaning of the relation (1.1) becomes more clear. 

Similarly, in the lepton sector, we define the eigenstates of 
family $\ell^0_\alpha = (\nu^0_\alpha, e^0_\alpha)$
 as follows:
$$
H_{int} = g_F (\overline{\ell^0})^\alpha \gamma_\mu  
\eqno(2.2)
$$
Here, the mixing matrices are given by
$$
\left( \begin{array}{c}
\nu_1 \\
\nu_2 \\
\nu_3
\end{array} \right) = U_\nu 
\left( \begin{array}{c}
\nu_1^0 \\
\nu_2^0  \\
\nu_3^0
\end{array} \right), \ \ \ \ 
\left( \begin{array}{c}
e_1 \\
e_2 \\
e_3
\end{array} \right)  \equiv
\left( \begin{array}{c}
e \\
\mu \\
\tau
\end{array} \right)
= U_e 
\left( \begin{array}{c}
e_1^0 \\
e_2^0  \\
e_3^0
\end{array} \right) ,
\eqno(2.3)
$$
where the SU(2)$_L$ doublets are given by
$$
\left( \begin{array}{c}
\nu^0 \\
e^0
\end{array} \right)_L = 
\left( \begin{array}{c}
U_\nu^\dagger \nu  \\
U_e^\dagger e
\end{array} \right)_L = U_e^\dagger
\left( \begin{array}{c}
U_e U_\nu^\dagger \nu \\
 e
\end{array} \right)_L .
\eqno(2.4)
$$
Here, $\nu_i=(\nu_1, \nu_2, \nu_3)$ and 
$e_i = (e, \mu, \tau)$ are eigenstates of mass.

As seen in Eq.(2.4), note that the SU(2)$_L$ partners 
$(\nu_e, \nu_\mu, \nu_\tau)_L$
of $e_L=(e, \mu, \tau)_L$ are 
$U_e U_\nu^\dagger\, \nu_L$.
  
%The purpose of the present paper is how to confirm
%whether the charged leptons are in the simultaneous 
%eigenstates for ``mass" and ``family" or not.

The neutrino family mixing, Maki-Nakagawa-Sakata 
(MNS) matrix $U_{MNS}$ \cite{MNS}, is defined by 
$$
\left( \begin{array}{c}
\nu_e \\
\nu_\mu \\
\nu_\tau
\end{array} \right) = U_{MNS} 
\left( \begin{array}{c}
\nu_1 \\
\nu_2  \\
\nu_3
\end{array} \right) ,
\eqno(2.5)
$$
i.e.
$$
\nu_\ell = (U_{MNS})_{\ell i} \nu_i ,   
\eqno(2.6)
$$
so that 
$$
U_{MNS} = U_e (U_\nu)^\dagger .
\eqno(2.7)
$$
Again, we find that we cannot see ``$U_e ={\bf 1}$ 
or not" from the observed neutrino data.

If it is  $U_e \neq {\bf 1}$,  
we will see the characteristic events 
$\mu$-$e$ conversions. 
In the next section (Sec.3), we will discuss a $\mu$-$e$ 
conversion
$$
\mu +N \rightarrow e + N . 
\eqno(2.8)
$$
(However, as we discuss in the next section, the $\mu$-$e$ 
conversion (2.8) will also take a place for the case 
$U_e = {\bf 1}$. )

In the section 3, as a more clear event for $U_e \neq {\bf 1}$, 
we will discuss a typical $\mu$-$e$ conversion
$$
\mu \rightarrow e +\gamma .
\eqno(2.9)
$$
The decay (2.9) usually happens when we introduce an explicit  
lepton family-number violation term. 
However, we would like to emphasize that if we have family 
gauge bosons, this decay (2.9) will take place even if 
we do not introduce such an exotic term. 
(For example, for earier work of $\mu \rightarrow e+\gamma$,
see Ref.\cite{PRL59}.)

\vspace{5mm}

{\large\bf 3. \ The $\mu$-$e$ conversion $\mu N \rightarrow e N$ 
due to family gauge bosons }

\vspace{2mm}

For simplicity, we neglect a mixing between $\mu$  and 
$\tau$, so that we consider only a mixing between $e$ and $\mu$: 
$$
\left( 
\begin{array}{c} 
e \\
\mu 
\end{array} 
\right) = \left( 
\begin{array}{cc} 
\cos\theta & \sin\theta  \\
-\sin\theta & \cos\theta 
\end{array} 
\right) \left(
\begin{array}{c} 
e^0 \\
\mu^0 
\end{array} 
\right) .
\eqno(3.1)
$$

If  $U_e \neq {\bf 1}$, the effective $\mu$-$e$ conversion 
term $\overline{e} \gamma_\rho \mu$ is induced as follows: 
$$
\overline{e} \gamma_\rho \mu = 
(\cos\theta \,  \overline{e^0}+ \sin\theta \, \overline{\mu^0})
\gamma_\rho ( -\sin\theta\, \mu^0 - \cos\theta\, \mu^0) 
$$
$$
= \cos^2\theta\, \overline{e^0} \gamma_\rho \mu^0 
-\sin^2\theta \, \overline{\mu^0} \gamma_\rho e^0 
+ \sin\theta \cos\theta\,  ( \overline{\mu^0}
\gamma_\rho \mu^0 - \overline{e^0} \gamma_\rho e^0) .
\eqno(3.2)
$$
The first and second terms interact with $A_1^{\ 2}$ and 
$A_2^{\ 1}$, respectively.
The third and forth terms interact with $A_2^{\ 2}$ and
$A_1^{\ 1}$, respectively.
However, hereafter, we will neglect the reactions via $A_2^{\ 1}$ 
and $A_2^{\ 2}$, because that via $A_2^{\ 1}$ has a small factor 
$\sin^2 \theta$ and that via $A_2^{\ 2}$ has a large mass 
$M_{22} \simeq \sqrt{2}M_{21} \sim 10^2$ TeV (see Eq.(3.9) later).

 \vspace{4mm}
 
%%%%%%%% Fig.1
\begin{figure}[h]
%\begin{figure}
\begin{center}
\includegraphics[clip, width=60mm]{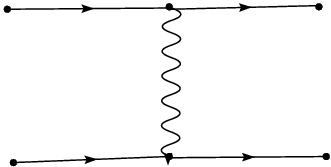}
\hspace{10mm}
\includegraphics[clip, width=60mm]{muN-eN.eps}
\end{center}

\vspace{-42mm}\hspace{12mm} 
$\mu = \cos\theta \, \mu_0 + \cdots$ \ \ \ 
$e = \cos\theta \, e_0 + \cdots$ \ \ \ \ \ 
$\mu = -\sin\theta \, e_0 + \cdots$ \ \ \ 
$e = \cos\theta \, e_0 + \cdots$

\vspace{14mm}\hspace{33mm} $A_2^{\ 1}$ 
\hspace{63mm} $A_1^{\ 1}$ 

\vspace{11mm}\hspace{12mm} 
$u = \cos\theta_u \, u_0 + \cdots$ \ \ \ 
$u = \sin\theta_u \, c_0 + \cdots$ \ \ \ \ \ \ \ 
$u = \sin\theta_u \, u_0 + \cdots$ \ \ \ 
$u = \sin\theta_u \, u_0 + \cdots$ 

\hspace{12mm}
$d = \cos\theta_d \, d_0 + \cdots$ \ \ \ 
$d = \sin\theta_d \, s_0 + \cdots$ \ \ \ \ \ \ \ 
$d = \sin\theta_d \, d_0 + \cdots$ \ \ \ 
$d = \sin\theta_d \, d_0 + \cdots$

\vspace{3mm}\hspace{45mm} (a) \hspace{60mm} (b)

\caption{ \ $\mu + N \rightarrow e+N$ via 
$A_2^{\ 1}$ and $A_1^{\ 1}$ }

\end{figure}
%%%%%%%%%%%%%%%%%%%%%%%%%%%%%%% end of Fig.1

The reaction $\mu N \rightarrow e N$ (2.7) can take a place
via the family gauge boson $A_2^{\ 1}$,  
even when   $U_e = {\bf 1}$, as shown in Fig.1 (a). 
When $U_e \neq {\bf 1}$, in addition to the diagram Fig.1 (a), 
the diagram Fig.1 (b) via $A_1^{\ 1}$  becomes possible. 

In order to image explicit masses of the family gauge bosons, . 
we adopt the Sumino's family gauge boson model
\cite{Sumino09},   where 
the family gauge boson masses are given by
$$
(M_{\alpha \beta})^2 = k_f (m_{ei} + m_{ej} ) , 
\ \ \ \ (\alpha \simeq i, \ \beta \simeq j) ,
\eqno(3.3)
$$
where $k_f$ is a constant with mass dimension. 
(Exactly speaking, Sumino has assumed only $(M_{ii})^2 \propto (m_{ei})^n$ 
where $n$ is free. 
However, in this paper, we take a more explicit form (3.3)
with $n=1$.   
If we assume a case with $n \geq 2$, our prediction in this 
section will be more enhanced visibly. )
Then, the mass ratio $(M_{11}^2/M_{12}^2)^2$ is given by
$$
\left( \frac{ M_{11}^2}{ M_{12}^2} \right)^2 = 
\left( \frac{ 2 m_e}{m_\mu+m_e} \right)^2 \sim (10^{-2})^2 \sim 10^{-4}  .
\ \ \ i.e. \ \ \ \frac{M_{11}}{M_{12}} \sim 10^{-1} . 
\eqno(3.4)
$$
Therefore the ratio of the transitions via $A_2^{\ 1}$ vs 
$A_1^{\ 1}$ is given by
$$
\frac{\sigma(A_1^{\ 1})}{\sigma(A_2^{\ 1}) } \simeq 
\left(\frac{\sin\theta \cos\theta}{\cos^2\theta}\right)^2
\left( \frac{ \cos^2 \theta_{u,d} }{ 
\sin\theta_{u,d} \cos\theta_{u,d} } \right)^2 
\left( \frac{M_{12} }{M_{11} }\right)^4 
$$
$$
\sim \left(\frac{\sin \theta}{\cos\theta} \right)^2
\left( \frac{\cos \theta_{u,d}}{\sin\theta_{u,d}}\right)^2 
\times 10^4 
\sim \theta^2 \times 10^5 ,    
\eqno(3.5)
$$ 
where we regarded as $\sin^2\theta_{u,d}\sim 10^{-1}$. 
(However, note that we know the value of the Cabibbo angle 
$\theta_C= \theta_d -\theta_d$, but we do not know the
values of $\theta_u$ and $\theta_d$ separately.)  
Then, if $\theta \geq 10^{-2}$, the case via  $A_1^{\ 1}$
(i.e. the case $U_e \neq {\bf 1}$) will be  
enhanced compared with the case via $A_2^{\ 1}$. 

(Thus, although the  numerical estimates are dependent on the
assumption (3.3), the results are only order-estimates, so that 
the choice (3.3) is not so essential.)
  
However, note that we cannot distinguish $\sigma(A_1^{\ 1})$ 
from  $\sigma(A_2^{\ 1})$ by means of the observation of
 the reaction  $\mu N \rightarrow e N$ because of the 
unknown parameters $\theta_u$ and $\theta_d$. 
Of course,  it is basically  possible that we nows the values of 
$\sin\theta_u$ and $\sin\theta_d$ separately when 
we analyze a nucleon dependency (for example, 
see Ref. \cite{K-Y_16}). 
But, the analysis will be not so easy, 
so that the study is practically impossible.    
 
Thus, our rough  prediction is given by
$$
Br (\mu N \rightarrow e N) \simeq 
\frac{\sigma(\mu N \rightarrow e N)}
{\sigma(\mu N \rightarrow \mu N)}
$$
$$
\sim \sin^2 \theta_{u,d} 
 \left(\frac{g_F}{g_Z} \right)^2 
\left( \frac{M_Z}{M_{12}} \right) ^4 
 \sim \left( \frac{M_Z}{M_{12}} \right)^4 
 \times 10^{-2},
\eqno(3.6)
$$
where we put $\sin^2 \theta_{u,d} \sim 10^{-2}$.

At present, the experimental observation limit \cite{SINDRUM}
is
$$
Br(\mu N \rightarrow e N) \equiv
\frac{\sigma(\mu N \rightarrow e N)}
{\sigma(\mu N \rightarrow all)} < 7 \times 10^{-13},
\eqno(3.7)
$$
for $N= Au$.

From Eqs.(3.6) and (3.7),  we obtain a constraint 
$$
\left( \frac{M_Z}{M_{12}} \right)^4 < 10^{-11} \ \Rightarrow 
\ M_{12}  >  M_Z \times 10^3 \sim 10^2 \ {\rm TeV} ,
\eqno(3.8)
$$
so that it will be impossible to observation to observe 
the family gauge bosons directly. 
(Exactly speaking, the predicted values are dependent on the nuclear 
target.  For example, see Ref.\cite{K-Y_16}.) 
 
 Hereafter, we suppose
 $$
 M_{21} \sim 10^2 \ {\rm TeV}, \ \ \ \ 
 M_{11} \sim 10^1 \ {\rm TeV},
\eqno(3.9)
$$
from Eq.(3.4) and considering of no observation of the family 
gauge bosons at LHC.

 In general, the adoption of a family gauge boson  model induces 
family-changing neutral currents, so that the adoption 
must be highly suppressed.  
In the present model with  the constraint (3.9),   
such neutral-current troubles in $K^0$-$\bar{K}^0$ mixing, and so on, 
are negligibly suppressed.

%%%%%%%%%%%%%%%%%%%%%%%%%%%%%%%%%%%%%%%%%%%%%%%%%%%%%
%\newpage

\vspace{5mm} 
.

{\large\bf 4. \ The radiative decay $\mu \rightarrow e + \gamma$} 

\vspace{2mm}

Let us show that for the case $U_e \neq {\bf 1}$ we can expect the 
observation of the  decay $\mu \rightarrow e + \gamma$ 
without introducing any explicit $\mu$-$e$ conversion term.

The decay amplitude ${\cal M}(\mu \rightarrow e + \gamma)$ 
due to a family gauge boson exchange is given by the 
three diagrams in Fig.2, 

%%%%%%%%%%%%%%%%%%%%%%%%%%%%%%%%%%%%%%%%%%%%%%%%%%%
%%%%%%%% Fig.2

%%%%%%%%%%%%%%%%%%%%%%%%%%%%%%%%%%
\begin{figure}[th]
%\begin{center}
%%%%%%%%%%%%%%%%%%%%%%%%
\vspace{10mm}
\hspace{5mm}
\includegraphics[clip, width=40mm]{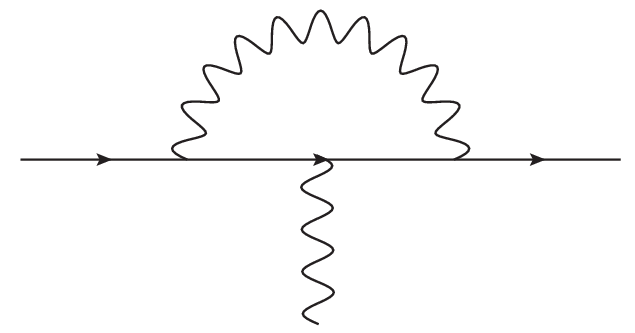}
\hspace{5mm}
%\vspace{-80mm}
%
\includegraphics[clip, width=40mm]{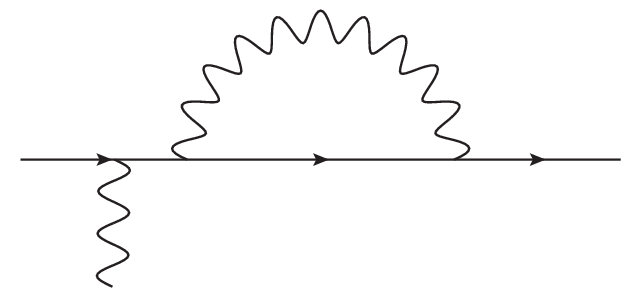}
\hspace{5mm}
\includegraphics[clip, width=40mm]{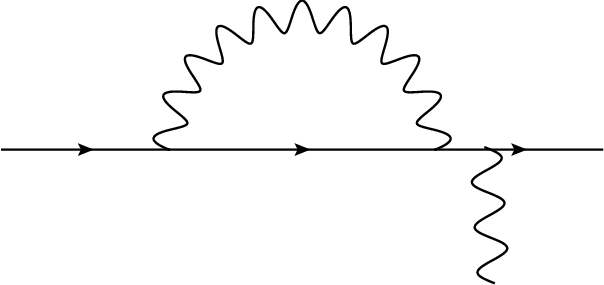}
%\end{center}

%%%%%%%
\vspace{-22mm}\hspace{10mm} $A_\alpha^{\ \beta}$ 
\hspace{16mm} $A_\beta^{\ \alpha}$ 

\vspace{4mm}\hspace{10mm}$e_2$ \hspace{4mm}$e_k$
\hspace{6mm}$e_k$ \hspace{4mm}$e_1$

\vspace{5mm} \hspace{24mm}$\gamma$

\vspace{3mm}\hspace{20mm} (a) \hspace{38mm} (b)\hspace{38mm} (c) 

%%%%%%
%%%%%%%%%%%%%%%%%%%%%%%%%%
\vspace{-10mm}
\hspace{30mm} \caption{ Three diagrams for $\mu \rightarrow e + \gamma$ }
\vspace{5mm}
\end{figure}
%%%%%%%%%%%%%%%%%%%%%%%%%%%%%%% 

The transition amplitude is given by 
$$
{\cal M}(\mu \rightarrow e + \gamma) = g_f^2 e \, 
\overline{u_1}(p) (U_e)_1^{\ \beta} ((U_e)^\dagger)_\alpha^{\ k} 
F_\rho (m_k, M_{\alpha\beta}) \, \varepsilon^\rho (p-p)
(U_e)_k^{\ \alpha} ((U_e)^\dagger)_\beta^{\ 2} u_2(q) ,
\eqno(4.1)
$$
where $\varepsilon_\rho$ is a polarization vector of photon, 
and $\overline{u_1}= \overline{u_e}$ and $u_2 =u_\mu$. 
The function $F_\rho (k_k, M_{\alpha\beta})$ is given by
$$
F_\rho (m_k, M_{\alpha\beta}) = 
F^{(a)}_\rho (m_k, M_{\alpha\beta}) +
F^{(b)}_\rho (m_k, M_{\alpha\beta}) +
F^{(c)}_\rho (m_k, M_{\alpha\beta}),
\eqno(4.2)
$$
correspondingly to the three diagrams in Fig.2. 
For example, $F^{(a)}$ is given  by
$$
 F^{(a)}_\rho (m_k, M_{\alpha\beta}) = - i^3 i^3 
 \int \frac{d^4 k}{(2\pi)^4} \gamma_\mu  
\frac{\not{p}-\not{k} + m_k}{(p-k)^2 -m_k^2} \gamma_\rho  
\frac{\not{q}-\not{k} + m_k}{(q-k)^2 -m_k^2} \gamma_\nu
\frac{g^{\mu\nu} }{k^2 -M_{\alpha\beta}^2 } . 
\eqno(4.3)
$$

%%%%%%%% Fig.3
\begin{figure}[h!]
%\begin{figure}
\begin{center}
\includegraphics[clip, width=60mm]{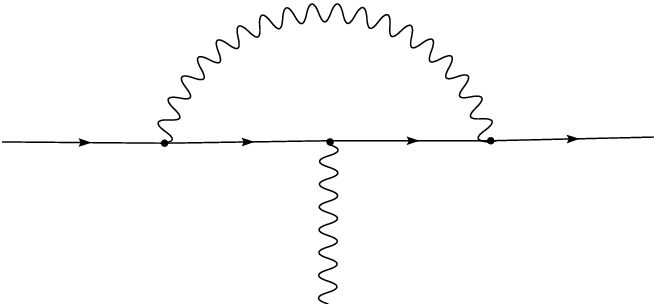}
\end{center}
%\vspace{-28mm}\hspace{55mm} $A_\alpha^{\ \beta}$ 
%\hspace{33mm} $A_\beta^{\ \alpha}$ 
\vspace{-32mm}\hspace{60mm}{$k$}

\vspace{5mm}\hspace{54mm}$q$ \hspace{10mm}$q-k$
\hspace{10mm}$p-k$ \hspace{10mm}$p$

\vspace{8mm} \hspace{75mm}$q-p$

\vspace{1mm}
\caption{ \ Fyenmann diagram corresponding to the diagram (a) 
in Fig. 2} 
\end{figure}
%%%%%%%%%%%%%%%%%%%%%%%%%%%%%%% 

%%%%%%%%%%%%%%%%%%%%%%%%%%%%%%%%%%%%%%%%%%%%%
\vspace{2mm}

Note that the charged lepton states which interact  
with $A_\alpha^{\ \beta}$ are $(e^0)_\alpha$, 
while the charged lepton states which 
propagate in the loop diagram are $e_k$.

In this section, for the purpose to see an actual example,
let us discuss a case with only $e\leftrightarrow \mu$ mixing. 
The mixing is approximately given by (3.1). 
Then, the decay amplitude is given by
$$
{\cal M}(\mu \rightarrow e +\gamma) = g_F^2 e 
\sin\theta \cos\theta \, \overline{u_e}(p) 
\left\{ \right.
$$
$$
 - \left[  \cos^2\theta\ F_\rho (m_e, M_{11}) \right. 
+\left. \sin^2 \theta\ F_\rho (m_\mu, M_{11}) \right]
$$
$$
+\left[ 
\right. \sin^2\theta\ F_\rho (m_e, M_{22}) 
\left. +\cos ^2 \theta\ F_\rho (m_\mu, M_{22})  
 \right] 
 $$
 $$
+(\cos^2 \theta -\sin^2\theta) \left[
 F_\rho (m_e, M_{12})   -  F_\rho (m_\mu, M_{12})  \\
 \right]\left. \right\} \, u_\mu (q)\, \varepsilon^\rho (q-p) .
\eqno(4.4)
$$

Note that the matrix element cannot become zero even if 
in the limit of $m_\mu=m_e$. 
On the other hand, the matrix element can become zero 
in the case that all family gauge boson masses are degenerated
even if in the case $m_\mu \neq m_e$. 
This means that if the mass differences among $M_{ij}$ 
are negligibly small, the amplitude with $m_k= m_e$ in  Eq..(4.4) 
becomes negligibly small. 

 Since the function  $F_\rho$ is dimensionless, in general, 
the value of $F_\rho$ is given by
$$
F(m_k, M_{\alpha. \beta} )_\rho \simeq \frac{1}{16\pi^2}
\frac{1}{M^2_{\alpha\beta}} 
\left( 
c_0 M^2_{\alpha\beta} + c_1 m_k^2 
\right) \gamma_\rho .
\eqno(4.5)
$$
Here, the coefficients $c_0$ and $c_1$ are
given by numerical values with the order  $O(1)$.
The coefficient $c_0$ is independent of the structure of 
$F(m_k, M_{\alpha\beta})$,
while $c_1$ is dependent on the structure of
 $F(m_k, M_{\alpha\beta})$.
 Therefore, from the expression (4.4),  we find that 
 the first term with the coefficient $c_0$ cannot 
 contribute to the ${\cal M}(\mu \rightarrow e +\gamma)$.
 Only the $c_1$ can contribute to the decay amplitude (4.4). 

Now, we estimate which term is dominant in Eq.(4.4).   
Here, according to Sumino's speculation, 
we use family gauge boson masses (3.3), so that 
we use the relation (3.4), $M_{11}/M_{12} \sim 10^{-1}$.  
Then, the dominant terms in Eq.(4.4) will be
terms with gauge boson masses $M_{11}$:
$$
{\cal M}(\mu \rightarrow e +\gamma) 
$$
$$
\simeq  \frac{g_F^2\, e}{16\pi^2}
 \sin\theta \cos\theta\,  \overline{u_e}(p)
 \left[  \cos^2\theta\ F_\rho (m_e, M_{11}) \right. 
+\left. \sin^2 \theta\ F_\rho (m_\mu, M_{11}) \right] 
u_\mu (q) \varepsilon^\rho (q-p) 
$$
$$
\simeq \frac{g_F^2 e}{16\pi^2} \sin \theta \cos\theta 
\left( \cos^2\theta \frac{m_e^2 }{M_{11}^2} 
+ \sin^2 \theta \frac{m_\mu^2}{M_{11}^2} \right) \,
\overline{u}(p)\not{\varepsilon}\, u(q) 
$$
$$
\simeq    \theta 
\left[ \left( \frac{m_e}{m_\mu} \right)^2 
+\theta^2 \right] \left( \frac{m_\mu}{M_{11}} \right)^2 
m_\mu \times 3.5 \times 10^{-4} . 
\eqno(4.6)
$$
Here, we have put $g_F^2 e/16 \pi^2 = 2 e^3 /16\pi^2=
3.52 \times 10^{-4}$, where we used the relation
$\alpha_{fam} = 2 \alpha_{em}$ in the family gauge boson 
model \cite{Sumino09}. 

The predicted value of $\Gamma(\mu \rightarrow e \gamma)$ is 
sensitive of the value $(m_\mu/M_{11})^4$. 
Therefore, taking  the condition Eq.(2.9) into consideration, 
we denote the value of  $m_\mu/M_{11}$ as 
$$
 \frac{m_\mu}{M_{11}}  \equiv \xi \times 10^{-4} . 
\eqno(4.7)
$$
Then, we can denote $\Gamma(\mu\rightarrow e \gamma) $
as 
$$
\Gamma(\mu\rightarrow e \gamma) \simeq
\frac{1}{16\pi^2} \left| {\cal M} \right|^2 \frac{1}{m_\mu}
\simeq
\theta^2 \left[ \left( \frac{m_e}{m_\mu} \right)^2 
+\theta^2 \right]^2 \xi^4 \times 2.4 \times 10^{-15}\, {\rm MeV} .
\eqno(4.8)
$$

Now,  we can predict $Br(\mu\rightarrow e \gamma)$ as
$$
Br(\mu\rightarrow e \gamma)\simeq
\theta^2 \left[  \theta^2  
+  \left(\frac{m_e}{m_\mu} \right)^2 
\right]^2 \xi^4 \times 0.80 \times 10 , 
\eqno(4.9)
$$
where we have used $\Gamma(\mu\rightarrow all) =
1/(2.197 \times 10^{-6} {\rm s}) = 
2.996 \times 10^{-16}$ MeV. 

%%%%%%%%%%%%%%%%%%%%%%%%%%%%%%%%%%%%

\vspace{5mm}

{\large\bf  5. \ Rough estimate of  the mixing angle $\theta$ } 

\vspace{2mm}

The present experimental limit  of $Br(\mu\rightarrow e \gamma)$
\cite{PDG18}  is 
$$
Br(\mu\rightarrow e \gamma)  < 4.2 \times 10^{-13} . 
\eqno(5.1)
$$
Therefore,  from (4.9), we obtain
$$
\theta^2 \left[  \theta^2  
+  \left(\frac{m_e}{m_\mu} \right)^2 
\right]^2 \xi^4 < 1.5 \times 10^{-14}  ,
\eqno(5.2)
$$
i.e.
$$
\theta \left[  \theta^2  
+  \left(\frac{m_e}{m_\mu} \right)^2
\right] \xi^2 < 1.22 \times 10^{-7}  .
\eqno(5.3)
$$
Note that under the present estimate, the numerical 
factor $c_1$ in the expression (3.5) was taken as
$c_1=1$ for convenience.
The factor $c_1$ is not sensitive to our 
rough estimate of $\theta$ as compared with 
the parameter $\xi$. 
 
For convenience, we put
$$
\theta= x \times 10^{-2}, \ \ \ 
a=\left(\frac{m_e}{m_\mu}\right)^2  \times 10^4 
= 0.236 ,
\ \ \ b= \frac{1}{\xi^2}\times 0.122  , 
\eqno(5.4)
$$
so that we obtain a constraint for $x$:
$$
y \equiv x^3 + a x -b < 0 .
\eqno(5.5)
$$
Since $dy/dx = 3 x^2 +  a > 0$, the function $y(x)=0$ 
has only one real solution $x_0$ (the other two are imaginary), 
so that the constraint (5.5) leads to
$$
\begin{array}{l}
  x < x_0= 0.35, \ \ {\rm i.e. } \ \  \theta < \theta_0 =  0.35 \times 10^{-2} 
\ \ (\xi =1) , \\
  x < x_0= 0.13, \ \ {\rm i.e. } \ \  \theta <  \theta_0  = 0.13  \times 10^{-2} 
\ \ (\xi =2)  .
\end{array}
\eqno(5.6)
$$
Thus, even if $U_e \neq {\bf 1}$, 
the value $\theta$ of the family mixing 
between $e$ and $\mu$ is 
considerably small compared with 
the Cabbibo mixing $\theta_C = 0.22$ in quarks.

%%%%%%%%%%%%%%%%%%%%%%%%%%%%%%%%%%%%

\vspace{5mm}

{\large\bf 6. \ Conclusion } 

\vspace{2mm}

We have discussed on the topic ``are the charged leptons
on the simultaneous eigenstates of mass and family?".
If the observed charged leptons are mixing states among 
the eigenstates of  ``family" (i.e. $U_e \neq {\bf 1}$), 
the effect of $U_e \neq {\bf 1}$ will be observed in 
the reaction $\mu N \rightarrow e N$ and the decay 
$\mu \rightarrow e +\gamma$. 
However, since the observation  $\mu N \rightarrow e N$ can 
be also caused in the case $U_e = {\bf 1}$, 
the judgment for `` $U_e ={\bf 1}$ or not" is difficult. 

On the other hand, the observation is $\mu \rightarrow e +\gamma$
is characteristic phenomenon in the case of $U_e \neq {\bf 1}$.  
As seen in Sec.3, the $e$-$\mu$ mixing angle $\theta$ 
will be considerably small, 
$$
 \theta  \lesssim 10^{-3} , 
\eqno(6.1)
$$
in comparison with the Cabibbo mixing  angle $\theta_C = 0.22$ 
in quarks.
However, note this result (6.1) does not always mean $U_e = {\bf 1}$ 
in the strict sense. 
We still have a possibility $U_e \neq {\bf 1}$ with 
$\theta \lesssim 10^{-3}$.

%%%%%%%%%%%%%%%%%%%%%  
%
%\newpage
\vspace{5mm} 
%
%\newpage

{\large\bf Acknowledgements} % acknowledgement

\vspace{2mm}

The  author  is grateful to Prof. M. Yamanaka 
for his enjoyable discussion. 
And, also, the author is grateful to  Prof. T. Yamashita 
for his helpful assistance in preparing this manuscript.  
This work is supported by JPS KAKENHI Grant 
number JP1903826.

\vspace{15mm}

%%%%%%%%%%%%%%%%%%%%%%%%%%%%%%%%%%%%%%%%%%%
\newpage

%%%%%%%%%%%%%%%%%%%%%%%%%%%%%%%%%%%%%%%%%%%%%%%%%%%%

%\vspace{15mm}
%
%%%%%%%%%%%%%%%%%%%%%%%%%%%%%%%
 
%%%%%%%%%%%%%%%%%%%%
%%%%%%%%%%%%%%%%%%%%%%%%%%%%%%%

\end{document}